\documentclass[a4paper,11pt]{article}
\usepackage{pos}
\usepackage{hyperref}
\usepackage{lineno}

\title{Heavy flavor spectroscopy studies at CMS}

\manuallySeparateAuthors
\author[a]{Feng Zhu}
\author[]{ and } 
\author[a]{Kai Yi}
\author{on behalf of the CMS Collaboration}

\affiliation[a]{School of Physics and Technology, Nanjing Normal University,\\
Wenyuan Road No. 1, Nanjing, China}

\emailAdd{feng.zhu@cern.ch}
\emailAdd{yik@fnal.gov}

\abstract{

The CMS Collaboration has performed many studies in the field of heavy flavor spectroscopy. In this 
report, recent studies on exotic resonances in proton-proton collisions at $\sqrt{s} = 13$ TeV at CMS are presented. 
For the exotic hadrons, these results include the first evidence for X(3872) in heavy-ion collisions
and three new structures in $J/\psi J/\psi$ mass spectrum.
Beside the exotic hadrons, we also found new decay channels in conventional beauty hadrons
, 
including $B^{0}_{s}\rightarrow\psi(2S)K^{0}_{S}$, $\Lambda_{b}^{0}\rightarrow J/\psi\Xi^{-}K^{+} $, $B^{0} \rightarrow  \psi(2S)K^{0}_{S}\pi^{+} \pi^{-}$, and $\Xi_{b}^{-} \rightarrow \psi(2S)\Xi^{-}$.
}

\FullConference{%
  12th Large Hadron Collider Physics Conference - LHCP2024\\
  03 June 2024\\
  Boston, United States
}

\begin{document}
\maketitle

\section{$X(3872)$ studies}
Since the discovery of the $X(3872)$ by the Belle experiment in 2003~\cite{Belle:X3872}, 
which opened a new chapter for exotic mesons, many experiments have made significant efforts to study these particles. 
The CMS experiment at the LHC has performed numerous important studies. 
Among these are the measurement of the $X(3872)$ production cross section,
where the $X(3872)$ was reconstructed via its decay to $J/\psi \pi^+\pi^-$ in proton-proton collisions at $\sqrt{s} = 7$ TeV, 
providing crucial data on its production mechanisms~\cite{CMS:2013fpt}. 
Furthermore, the CMS experiment observed the $B^{0}_{s} \to X(3872)\phi$ decay using a data sample from proton-proton collisions at $\sqrt{s} = 13$ TeV, where the $X(3872)$ and $\phi$ are reconstructed in the $J/\psi \pi^+\pi^-$ and $K^+K^-$ channels respectively.
The decay mode provides new insights into the production dynamics and internal structure of the $X(3872)$~\cite{CMS:2020eiw}.

The $X(3872)$ was first observed by the Belle Collaboration~\cite{Belle:2003nnu}. 
Although its quantum numbers have been determined to be $J^{PC} = 1^{++}$ by the LHCb collaboration~\cite{LHCb:2013kgk},
its nature is still not fully understood. The production and survival of the $X(3872)$ in 
relativistic heavy ion collisions is expected to depend on the $X(3872)$'s internal structure~\cite{Zhang:2020dwn, Wu:2020zbx}.
Therefore, study of the $X(3872)$ production in relativistic heavy-ion collisions provides 
new opportunities to probe the nature of the $X(3872)$.

The CMS Collaboration has conducted a study on the production of the exotic state $X(3872)$ in lead-lead (Pb-Pb) 
collisions at a center-of-mass energy of $\sqrt{S_{NN}} = 5.02$ TeV, 
utilizing a sample collected in 2018 with an integrated luminosity of 1.7 nb$^{-1}$~\cite{CMS:2021znk}. 
This study marks the first evidence of $X(3872)$ production in such high-energy heavy ion collisions, 
with a statistical significance of 4.2 standard deviations.

\begin{figure}[!htbp]
  \centering
  \includegraphics[width=0.6\textwidth]{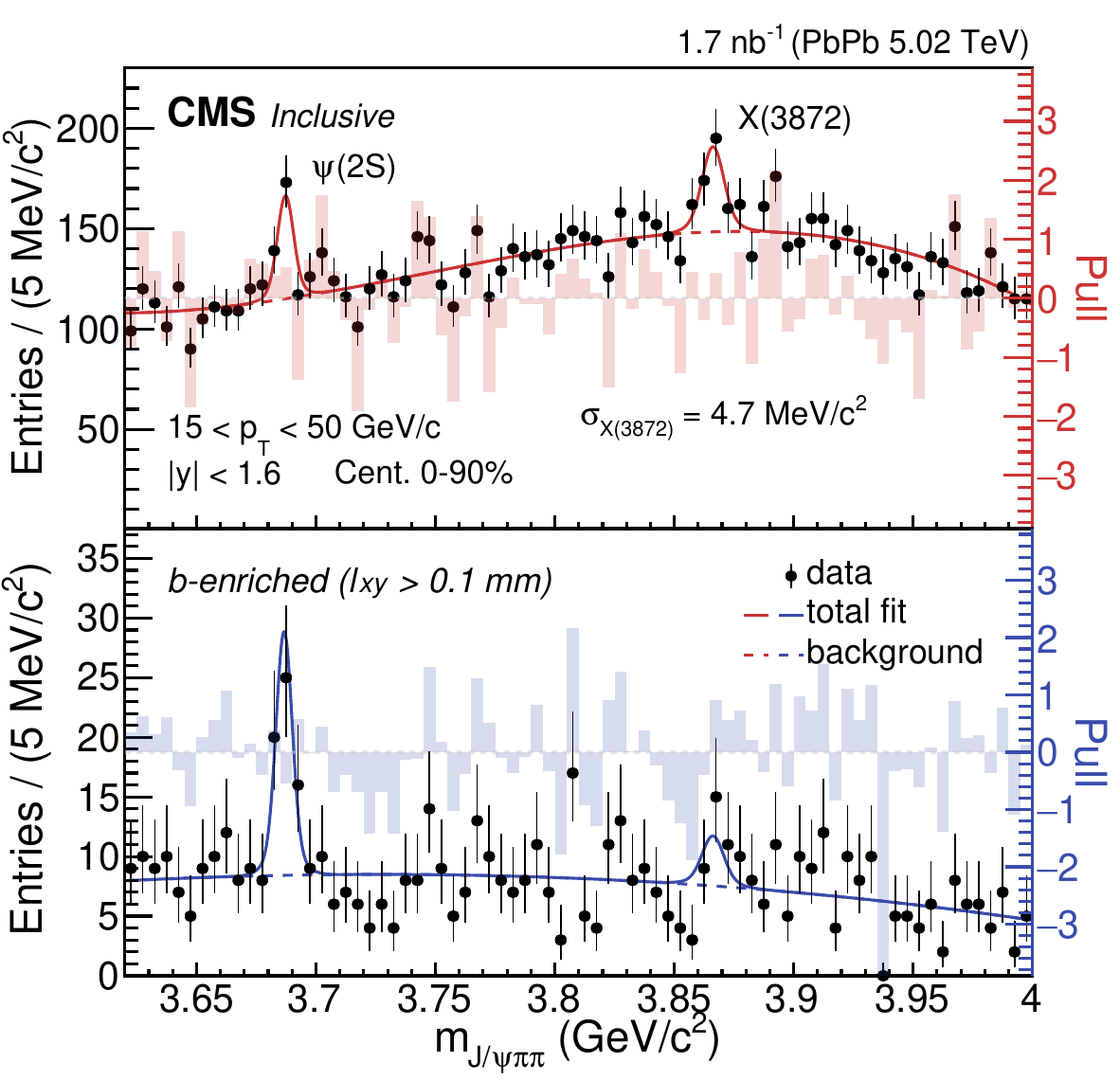}
  \caption{Invariant mass distribution of $m_{\mu\mu\pi\pi}$ 
  in Pb-Pb collisions,
  for the inclusive (upper) and $b$-enriched (bottom) samples~\cite{CMS:2021znk}. 
  The vertical lines on points represent statistical 
  uncertainties in the data. The results of the unbinned maximum-likelihood
  fits for the signal + background, and background alone, are also shown by the solid and dashed lines, 
  respectively. The pull distribution is represented by the shaded bars. The $X(3872)$ peak mass resolution,
  $\sigma_{X(3872)}$, calculated at the half-maximum of the signal peak, is also listed
  for reference.}
  \label{fig:x3872pbpb}
\end{figure}

The $X(3872)$ candidates are reconstructed through their decays into $J/\psi\pi^{+}\pi^{-}$, 
where the $J/\psi$ subsequently decays into $\mu^{+}\mu^{-}$. 
The candidates for the $X(3872)$ and $\psi(2S)$ states, a reference channel commonly used in such studies, are identified by the invariant mass distribution $m_{\mu\mu\pi\pi}$. 
Figure~\ref{fig:x3872pbpb} presents the observed invariant mass distribution for both $X(3872)$ and $\psi(2S)$ candidates: 
the upper plot illustrates the inclusive sample, 
while the lower plot displays the $b$-enriched sample, which is dominated by nonprompt candidates (transverse decay length $l_{xy} > 0.1$ mm).

The prompt production of $X(3872)$ was studied in the rapidity and transverse momentum ranges $|y| < 1.6$ and $15 < p_T < 50$ GeV/c, respectively. 
After subtracting the nonprompt contributions, which are primarily from $b$ hadron decays, 
the prompt yield ratio of $X(3872)$ to $\psi(2S)$ in Pb-Pb collisions was found to be $1.08 \pm 0.49 (\text{stat}) \pm 0.52 (\text{syst})$~\cite{CMS:2021znk}. 
This ratio is notably higher compared to typical values around 0.1 observed in proton-proton (pp) collisions at similar energies.

The production and suppression mechanisms of $X(3872)$ in the quark-gluon plasma (QGP) formed in heavy ion collisions provide 
critical insights into its internal structure. The observed enhancement of $X(3872)$ production in Pb-Pb collisions suggests 
possible coalescence mechanisms that favor the formation of this state in the QGP. 
This result serves as valuable input for theoretical models aiming to understand the nature and production mechanisms of exotic states 
like $X(3872)$~\cite{CMS:2021znk}.

\section{Observation of new structures in the $J/\psi J/\psi$ mass spectrum in pp collisions}
The $X(3872)$ and many other exotic candidates contain two heavy quarks ($c\bar{c}$). 
An analogue to heavy quarkonia would be fully heavy tetraquarks, which have been explored in theoretical models and 
are expected to be experimentally observable. The recent observation of the $X(6900)$ 
decaying into $J/\psi J/\psi$ has been reported by the LHCb Collaboration~\cite{LHCb:2020bwg}, 
and it was subsequently confirmed by the CMS and ATLAS experiments. 
In addition, the CMS
Collaboration reported the discovery of a new structure, X(6600), with a local significance of 7.9
standard deviations, along with evidence for another new structure provisionally named X(7100),
exhibiting a local significance of 4.7 standard deviations~\cite{CMS:2023owd}.

\begin{figure}[!htbp]
  \centering
  \includegraphics[width=0.95\textwidth]{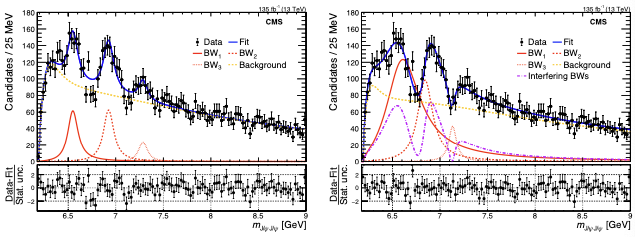}
  \caption{The fits to the $J/\psi J/\psi$ invariant mass spectrum in the CMS experiment. 
  The fit model incorporates three signal functions (X(6600)[BW1], X(6900)[BW2], and X(7100)[BW3]) 
  along with a background model (NRSPS+DPS+BW0). The fitting outcomes are presented in the left plot without considering interference effect, 
  and in the right plot with interference included~\cite{CMS:2023owd}.}
  \label{fig:jpsijpsi}
\end{figure}

The CMS experiment performed a study of the low-mass region of the $J/\psi J/\psi$ mass spectrum 
in pp collisions, using a data sample  corresponding to 
an integrated luminosity of 135 fb$^{-1}$ at a center-of-mass energy of 13 TeV~\cite{CMS-PAS-BPH-21-003}.
The two $J/\psi$ candidates are reconstructed using their $\mu^{+}\mu^{-}$ mode, 
and the final $J/\psi J/\psi$ mass distribution is shown in Fig.~\ref{fig:jpsijpsi}.
The measured masses and widths of three structures are summarized in Table~\ref{tab:jpsijpsi}.

\begin{table}[!htbp]
  \centering
   \begin{tabular}{cccc}
     \hline
     & BW1 & BW2 & BW3 \\
     \hline
     $m$ & $6552\pm10\pm12$ & $6927\pm9\pm5$ & $7287\pm19\pm5$ \\
     $\Gamma$ & $124\pm29\pm34$ & $122\pm22\pm19$ & $95\pm46\pm20$ \\
     $N$ & $474\pm113$ & $492\pm75$ & $156\pm56$ \\
     \hline
   \end{tabular}
   \caption{Summary of the fit results of the CMS $m(J/\psi J/\psi)$ distribution: the mass 
   $m$ and natural width $\Gamma$, in MeV, and the signal yields $N$ are given for three 
   signal structures~\cite{CMS-PAS-BPH-21-003}. The first uncertainties are statistical and the second systematic.}
   \label{tab:jpsijpsi}
\end{table}

Our $X(6900)$ parameters are in a good agreement with LHCb's non-interference result,
while the $X(6600)$ and $X(7300)$ are new structures.
In order to remove 
potential model dependencies in a comparison of the $X(6900)$ results, we also apply the principal 
two LHCb fit models to the CMS data, but using CMS-specific background shapes.
LHCb's Model I consists of the $X(6900)$ signal, NRSPS, non-resonant double parton scattering (NRDPS) and two more BWs -- 
around 6300 (BW0) and 6500 MeV (BW1) -- to account for the threshold enhancement.
LHCb's Model II consists of the $X(6900)$ signal, NRDPS, 
and the interference contribution of a Breit-Wigner structure $X(6700)$ and NRSPS.

In both models, the $X(6900)$ parameters are in a good agreement with LHCb's measurements, while our $X(6700)$ in Model II has 
a much larger amplitude and width compared to the LHCb's interfering Breit-Wigner, and none of the LHCb models provide a satisfactory description of our data.


It is interesting to highlight the two dips around 6750 MeV and 7150 MeV. Interference models are developed to
elucidate these dips. 
After investigating potential interference among the different components, 
the primary interference fit model is determined to involve the interference between 
three resonances $X(6600)[\text{BW1}]$, $X(6900)[\text{BW2}]$, and $X(7100)[\text{BW3}]$, 
implemented with a term proportional to $|r_1 \exp(i\phi_1) \text{BW1} + \text{BW2} + r_3 \exp(i\phi_3) \text{BW3}|$,
 where $r_{1,3}$ and $\phi_{1,3}$ denote the relative magnitudes and phases of BW1,3 with respect to BW2. 
 The fit outcome is illustrated in Fig.\ref{fig:jpsijpsi} (right), demonstrating an improvement in the $\chi^2$ probability in the signal region [6.2, 7.8] GeV to 65\%, 
 compared to only 9\% in the no-interference fit.

\section{Observations of new decay channels in conventional beauty hadrons}

Multibody decays of the $B$ mesons are well suited to the search for, and  study  of, exotic resonances. For example,
the discovery of $X(3872)$ was in $B\rightarrow K J/\psi\pi\pi$ decay~\cite{Belle:2003nnu}, and that of 
the first charged tetraquark candidate, $Z(4430)^{+}$, was in $B\rightarrow\psi(2S)K\pi^{\pm}$~\cite{Belle:2007hrb}.

The CMS experiment performed the first observation of the $B^{0}_{s}\rightarrow \psi(2S)K^{0}_{S}$ 
and $B^{0}\rightarrow\psi(2S)K^{0}_{S}\pi^{+}\pi^{-}$ decays, using a data sample
of proton-proton  collisions at $\sqrt{s} = 13$ TeV,
and an integrated luminosity of 103 fb$^{-1}$,
collected  in 2017 and 2018 ~\cite{CMS:2022cot}.
The $\psi(2S)$ and $K^{0}_{S}$ mesons are reconstructed using their decays into $\mu^{+}\mu^{-}$ 
and $\pi^{+}\pi^{-}$, respectively. 
Using the $B^{0}\rightarrow\psi(2S)K^{0}_{S}$ as a reference channel,
the relative branching fractions of 
$B^{0}_{s}\rightarrow\psi(2S)K^{0}_{S}$ and $B^{0}\rightarrow\psi(2S)K^{0}_{S}\pi^{+}\pi^{-}$ decays
are measured to be 
$\mathcal{B}(B^{0}_{s}\rightarrow\psi(2S)K^{0}_{S})/\mathcal{B}(B^{0}\rightarrow\psi(2S)K^{0}_{S})
=(3.33\pm 0.69(stat) \pm 0.11(syst)\pm 0.34(f_{s}/f_{d}))\times 10^{-2}$,
and $\mathcal{B}(B^{0}\rightarrow\psi(2S)K^{0}_{S}\pi^{+}\pi^{-})/
\mathcal{B}(B^{0}\rightarrow\psi(2S)K^{0}_{S}) = 0.480\pm 0.013 (stat) \pm 0.032 (syst)$,
where the last uncertainty in the first ratio corresponds to the uncertainty in the ratio of the production 
cross sections of $B^{0}_{s}$ and $B^{0}$ mesons.
With the currently available, statistics-limited data, 
the 2- and 3- body invariant mass distributions of the 
$B^{0}\rightarrow\psi(2S)K^{0}_{S}\pi^{+}\pi^{-}$ 
decay products do not show significant exotic narrow
structures in addition to the known light meson resonances.

\begin{figure}[!htbp]
  \centering
  \includegraphics[width=0.49\textwidth]{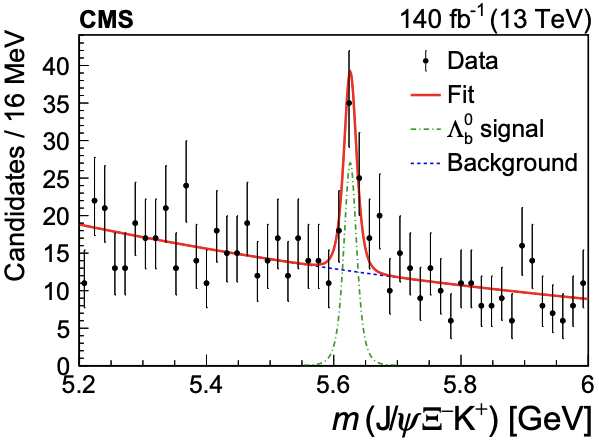}
  \includegraphics[width=0.49\textwidth]{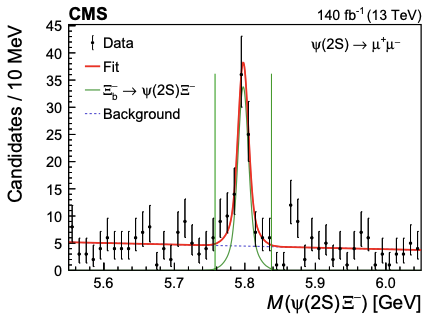}
  \caption{Measured mass distribution of  $J/\psi \Xi^{-}K^{+}$ (left) \cite{CMS:2024vnm}  and $\psi(2S)\Xi^{-}$
  (right) candidates~\cite{CMS:2022cot}.}
  \label{fig:newdecay}
\end{figure}

The CMS collaboration has also observed the decay of $\Lambda_b^0$ baryons into the final state $J/\psi \Xi^{-} K^{+}$ which is shown in the left of Fig~\ref{fig:newdecay}~\cite{CMS:2024vnm}. 
This decay channel is of particular interest as it can proceed through exotic intermediate resonances, 
such as pentaquark states. 
The decay $J/\psi p $  structure in  $\Lambda_b^0 \to  J/\psi p K^{-}$  observed by LHCb highlights the potential for unveiling new pentaquark states in the $\Lambda_b^0 \to  J/\psi \Xi^{-} K^{+}$ final state~\cite{LHCb:2015yax}. 
This observation is significant for the study of doubly-strange pentaquarks, providing new avenues for exploring the properties and interactions of these exotic states.


One of the major highlights is the first observation of the decay process $\Xi_b^{-} \rightarrow \psi(2S) \Xi^{-}$~\cite{CMS:2024rbi}. The observed invariant 
mass distribution of selected $\psi(2S) \Xi^{-}$ with $\psi(2S) \to \mu^{+} \mu^{-}$ candidates is shown in right of Fig~\ref{fig:newdecay}. 
This discovery is pivotal as it expands the understanding of baryon decays involving charmonium states. 
The $\Xi_b^{-}$ baryon, containing a bottom quark, decays into a $\Xi^{-}$ baryon and a $\psi(2S)$ meson, the latter being an excited state of the $J/\psi$ particle. 
The data used for this observation was collected from the full Run 2 of the LHC, amounting to an integrated luminosity of 140 fb$^{-1}$.

Additionally, the paper reports novel measurements of $b$-baryon properties, providing deeper insights into their characteristics and behavior. 
These measurements are essential for testing and refining theoretical models in quantum chromodynamics (QCD), the theory describing the strong interaction within particles containing quarks. 

The studies also include detailed analyses of the $\Xi_b^{*0}$ baryon, an excited state of the $\Xi_b^{0}$ baryon, whose mass and natural width are found to be $5952.4 \pm 0.1  (\text{stat + syst}) \pm 0.6  (m_{\Xi_b^{-}})$ MeV and $0.87^{+0.22}_{-0.20} (\text{stat}) \pm 0.16 (\text{syst})$ MeV, respectively~\cite{CMS:2024rbi}. 
These analyses contribute to the broader effort of mapping out the spectrum of heavy baryons and understanding their internal structure. 
Such studies are crucial for developing a comprehensive picture of hadronic matter and the forces that govern its interactions.

\section{Summary}
In summary, the CMS Collaboration has performed many studies in the field of hadron spectroscopy and exotic resonances. In this 
report, recent studies on exotic resonances in proton-proton collisions at $\sqrt{s} = 13$ TeV at CMS are presented. These results include
the first evidence for the $X(3872)$ in heavy ion collisions, observation of 
$B^{0}_{s}\rightarrow \psi(2S)K^{0}_{S}$, $\Lambda_{b}^{0}\rightarrow J/\psi\Xi^{-}K^{+} $, $B^{0} \rightarrow  \psi(2S)K^{0}_{S}\pi^{+} \pi^{-}$, and $\Xi_{b}^{-} \rightarrow \psi(2S)\Xi^{-}$  decays, 
and observation of new structures in the $J/\psi J/\psi$ mass 
spectrum.



\bibliographystyle{JHEP}
\bibliography{myrefs}

\end{document}